# The basis for design and manufacture of a DSP-based coincidence spectrometer

N. X. Hai[1], P. N. Tuan[1], N. N. Dien[1], D. Lanh[1], T. T. T. Huong[1], P. N. Son[1], P. D. Khang[2]

*[1]Nuclear Research Institute, 01 Nguyen Tu Luc, Dalat, Vietnam*

*[2] Nuclear Training Center, 140 Nguyen Tuan, Hanoi, Vietnam*

**Abstract:** The coincidence technique and the coincidence spectroscopy have been developed and applied for over 40 years. Most of popular coincidence measurement systems were based on analog electronics techniques such as time to amplitude conversion (TAC) or logic selecting coincidence unit. The above-mentioned systems are relatively cumbersome and complicated to use. With the strong growth of digital electronics techniques and computational science, the coincidence measurement systems will be constructed simpler but more efficient with the sake of application. This article presents the design principle and signal processing of a simple two-channel coincidence system by a technique of Digital Signal Processing (DSP) using Field Programmable Gate Arrays (FPGA) devices at Nuclear Research Institute (NRI), Dalat.

## I. Introduction

The background reduction measurement systems with active methods are based on coincidence or anti-coincidence techniques. These are mainly built from proper functional electronics modules in NIM or CAMAC standards [6, 8]. They allow us to identify coincidence or anti-coincidence events via main electronics blocks called coincidence unit or time to amplitude converter. Normally, basic configuration of a coincidence measurement system will, at least, consist of two channels, and the selection of coincidence event pairs depends on the defining moment of the pulses appearing at 'Timing' output of the pre-amplifier. Obvious drawback of this system is cumbersome in size, adjusting operation and synchronizing signals among electronics stages.

---

[1] *Corresponding author: Email address: nxhai@hcm.vnn.vn*

The growth of computer engineering and programmable devices that are capable of operating at high frequencies has allowed us to design a new spectrometry generation. The spectrometry generation is compact on size, simple in terms of connectivity and using [4, 7].

At Nuclear Research Institute (NRI), Dalat, a number of researches and construction of Compton suppression as well as event-event coincidence systems were performed in 1990s. The results of these studies were reported in ref. [3]. In the recent period at NRI, further studies on the coincidence spectrometer were presented in several publications [1, 2, 5]. Overall, although there have been significant improvements on acquiring as well as processing data, this system is still based on the traditional way in obtaining signals under the operation of a coincidence unit or TAC. With the research results gathered during the installation, investigation and exploitation of an 'event-event' coincidence spectrometer on digital signal processing (DSP) at NRI, we have now proposed a new approach for a design of the DSP-based multi-application coincidence technique through Field Programmable Gate Arrays (FPGA) devices. The basis for the design of this spectrometer will be presented within the framework of this article.

**2. The basis and method for the design**

*2.1. Design multi-channel analyzer*

The block diagram of MCA is presented in Figure 1. The functions of the diagram can be summarized as follows:

The pre-filter works as high-pass filter giving an output signal with a constant shaping time. In addition, the pre-filter has amplification function to generate the appropriate signals for ADC conversion.

The pre-filter' output signals are sent to flash ADC for sampling. The ADC' output signals are digital copies of input' analog signals.

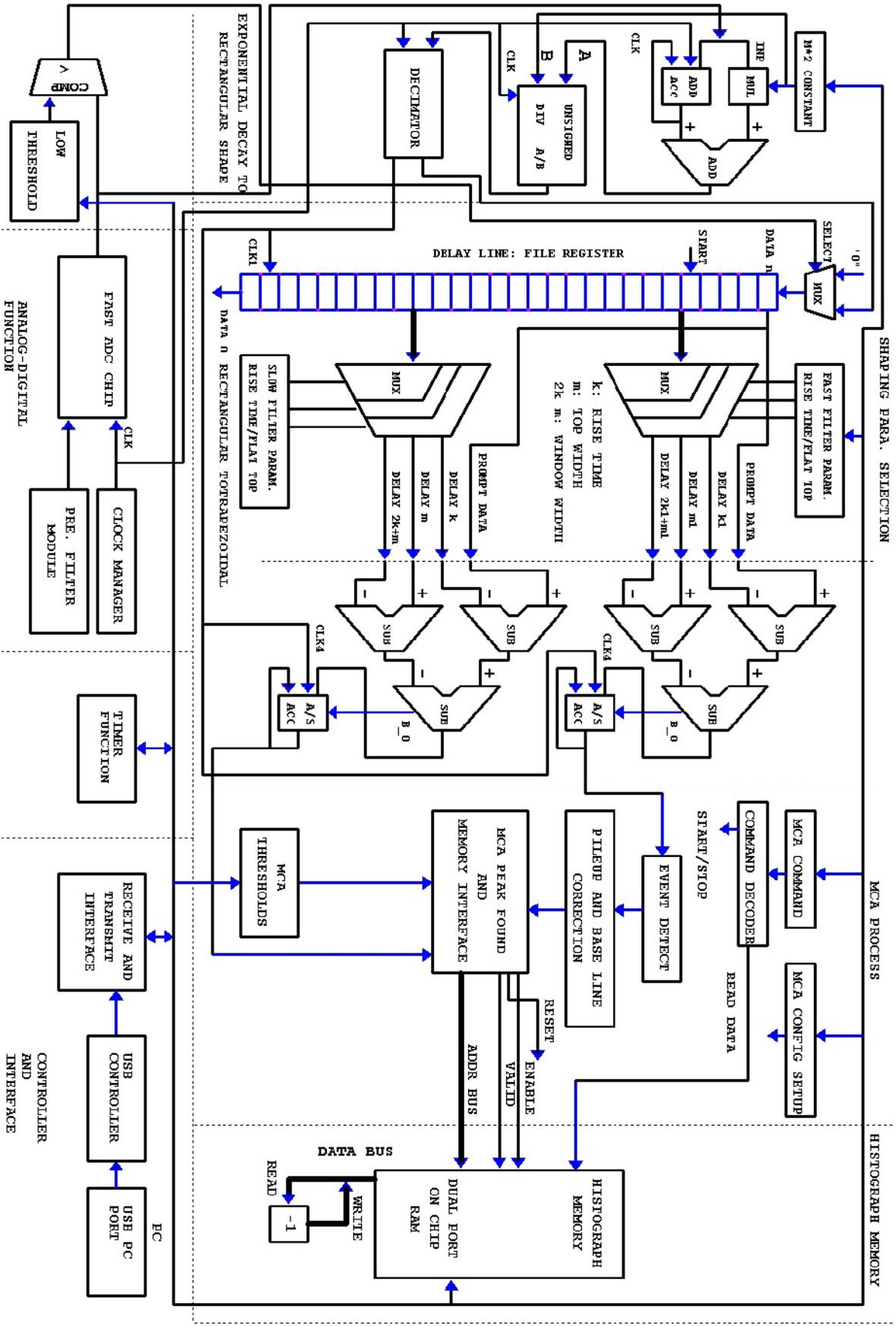

**Figure 1.** The MCA block diagram.

The transfering 'Exponential Decay to Rectangular Shape' circuit converts exponential decay to rectangular pulse, and its function is the opposite of high pass deconvoler (HPD). The transfering 'Rectangular Shape to Trapezoidal Shape' circuit converts shape of rectangular to trapezoidal pulse, and its function is the low pass filter (LPF). The circuits of pulse pile-up rejection, base line restoration and built-in configuration are also designed in the main board.

The pulse pile-up rejection circuit detects the pile-up pulses in duration from rise time to half of the flat width, in slow channel. In case of non-overlapped pulse is detected in monitoring duration, the right pulse will be analysis. A dual RAM port is integrated in FPGA device for buffering data. The preset time is set by user with capacity up to $4.2 \times 10^9$ seconds. The interface circuit is integrated in the main board and connected to PC through a USB-RS232 bridge.

*2.2. Design of the Event-Event coincidence spectrometer*

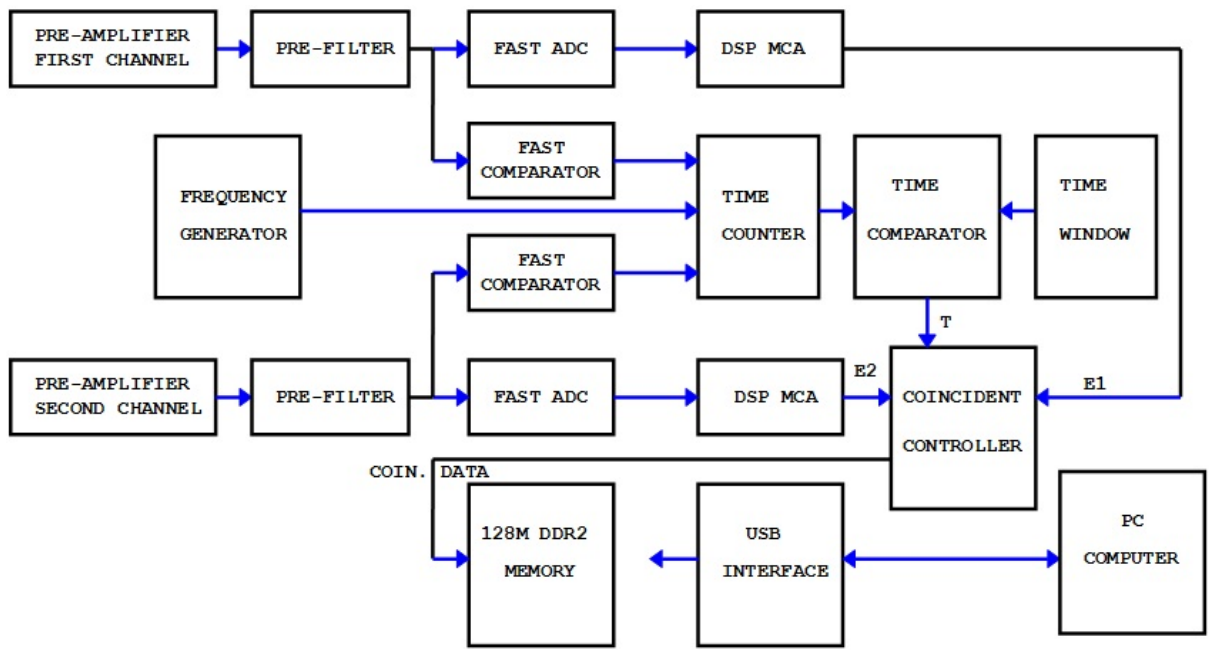

**Figure 2.** The block diagram of "event-event" coincidence spectrometer.

The "Exponential Decay to Rectangular Shape" circuit input' pulses shape of exponential decay to pulse shape of rectangular, this circuit function is opposite high pass deconvoler (HPD).

The "Rectangular Shape to trapezoidal shape" circuit input' pulse shape of rectangular to trapezoidal shape pulse, this circuit function is the low pass filter (LPF).

The pulse pile-up rejection, base line restorer and configuration circuits detect new events, pile-up pulses, level comparison, and base line restoration.

The pulse pile-up rejection circuit detects the pile-up pulses during time from rise time to half of the flat width of pulse in slow channel. If there is not any overlapped pulse in monitoring period, the pulse will be analysis. Memory for spectrum: the memory is a dual RAM port (which is) integrated in FPGA device. Preset time circuit: the preset time is set from data acquisition program up to a capacity of $4.2 \times 10^9$ seconds.

The interface circuit is integrated on the main board and connected to PC through USB-RS232 bridge.

*Principal operation:*

Two MCAs receive pulses from the preamplifier's energy outputs, the pulse' amplitudes are analysed (by DSP techniques); simultaneously, the preamplifier's output pulses are connected to fast comparator circuits. The output signals of these comparator circuits are used to start/stop a time counter. The used clock frequency is 400 MHz (2.5 ns period). The contents of the counter (time interval between two events) are compared with values of preset time interval), if the measured time is in the selected range, the amplitudes of pair of events will be recorded. Three values E1, E2 and T will be stored into memory; in which E1 and E2 are energy values of pair of coincidence gamma rays, and T is time interval. The measured results are read by data acquisition program and saved to file on the hard disk.

The PRE-FILTER, FAST ADC and DSP MCA are components of amplitude analysis channel. The FAST COMPARATOR, TIME COUNTER, TIME COMPARATOR, COINCIDENT CONTROLLER are components of coincidence channel. The used memory is DDR2 type.

## 3. Fabricating

*3.1 Hardware*

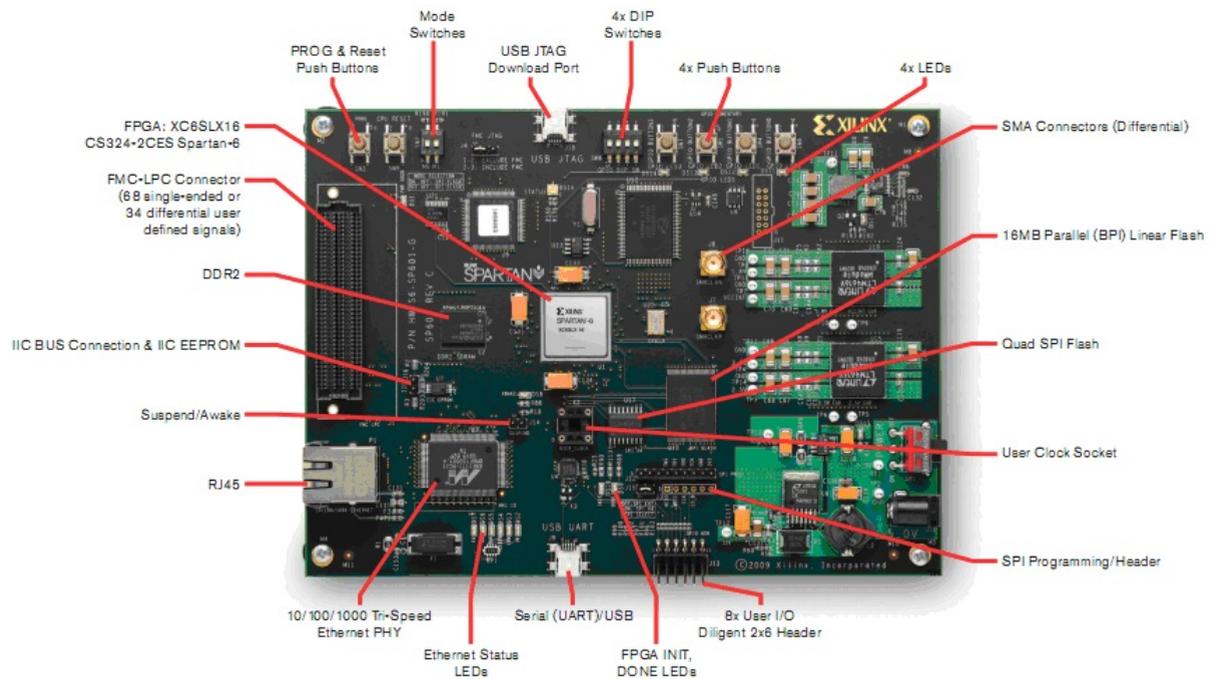

**Figure 3.** The main board.

The entire design is programmed and configured in the FPGA chip XC6LX16-CS324. The used main board is SP601 kit supplied by Xilinx as Figure 3.

*3.2. Data acquisition program*

The data acquisition program for energy and timing spectra was written under LabWIEW 8.5. The functions of program are as follows:

+ Connecting peripheral devices to PC and interfacing to PC through USB port.

+ Control of data acquisition for multi-MCA mode: start/stop data acquisition, preset measurement time, data saving, …

+ Control of data acquisition for "event-event" coincidence mode: start/stop data acquisition, preset measurement time, data saving, …

+ The base data analysis: display spectra, counts per channel, energy calibration, zoom in and zoom out spectra.

*3.3. The main features of spectroscopy*

1. The multi-channel analysis:

- Input: accept the preamplifier's output pulses from semiconductor detector, the amplitude from tens to 500 mV;

- Resolution: 8192 channels;

- Capacity counts / channel: $2^{32}$ -1;

- Preset time capacity: $2^{32}$-1 seconds with 1 second step;

- The integral nonlinearity: $< \pm 0.05\%$;

- Shaping time: 0.4; 0.8; 1.6; 3.2; 6.4; 6.12; 2.25 μs;

- Coarse gain: 1, 2, 5 and 10;

- Fine gain: $0.75 \div 1.25$, step 0.01;

- Displaying and rejecting the pile-up pulses, baseline correction and deadtime correction.

2. Event-event coincidence:

- Preset coincidence time window: $5 \div 10\,000$ ns, resolution step 5 ns;

- Mode: coincidence, PHA;

- Memory capacity: 16M events.

3. Power supply:

+ 12V/800mA,

- 12V/50mA,

+ 6V/300mA,

- 6V/100mA.

4. Dimensions: width NIM 2M.

**3. Testing and results**

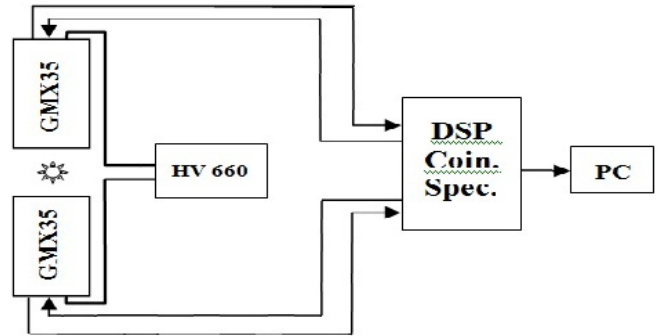

**Figure 4.** Experiment configuration.

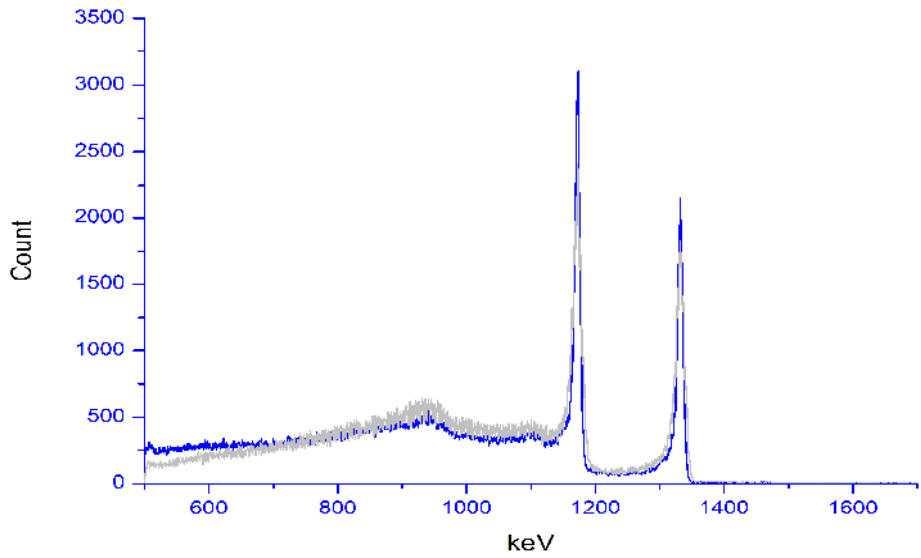

**Figure 5.** Energy spectrum of $^{60}$Co.

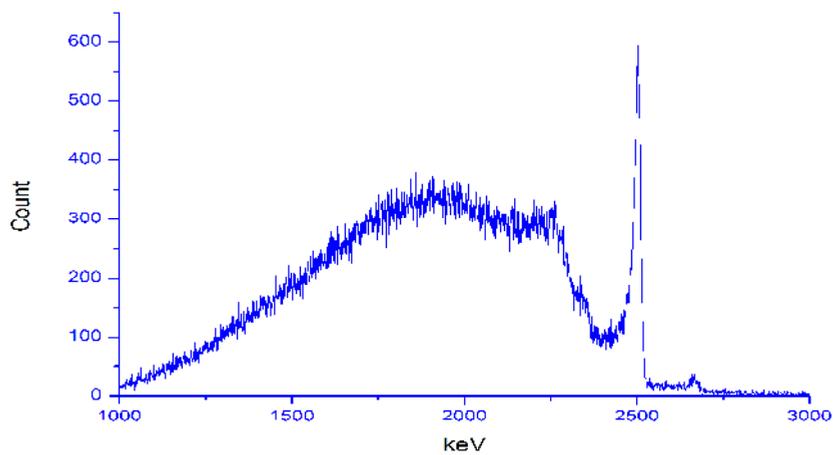

**Figure 6.** Summation spectra of $^{60}$Co.

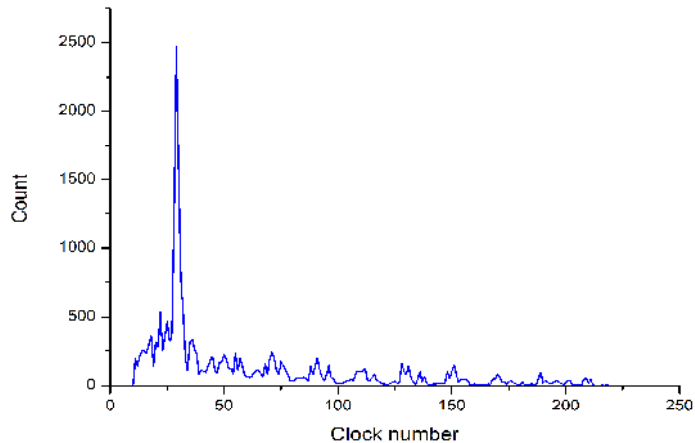

**Figure 7.** Timing spectra (one clock unit equivalent 5 ns).

## 4. Conclusions

The design of systems based on DSP techniques using FPGA allows to construct a simple, compact and impact coincidence spectrometer in which all of parameters are selected and controlled by software.

Currently, the diagram in Fig. 2 has been studying and constructing at the Department of Nuclear Physics and Electronics, NRI, Dalat, Vietnam. It is hopeful that, in the near future, the design of system might be the basis for development and application of coincidence measurement techniques in the field of physics research and applications.